\documentclass[twocolumn,showpacs,preprintnumbers,amsmath,amssymb,prl]{revtex4}

\usepackage{graphicx}
\usepackage{dcolumn}
\usepackage{bm}
\usepackage{lipsum}
\usepackage{amsmath}
\usepackage{relsize}
\usepackage{mathrsfs}
\usepackage{pgffor}
\usepackage[all]{xy}
\usepackage{color}
\usepackage{enumerate}
\usepackage{epstopdf}
\usepackage{color}
\usepackage{hyperref}

\begin{document}

\preprint{}

\title{
Tuning High-Harmonic Generation by Controlled Deposition of Ultrathin Ionic Layers on Metal Surfaces
}

\author{N\'{e}stor F. Aguirre$^{1}$}
\author{Fernando Mart\'{i}n$^{1,2,3}$}

\affiliation{
\vspace{0.3cm}
$^1$ Departamento de Qu\'{\i}mica, M\'odulo 13, Universidad Aut\'onoma de Madrid, 28049, Madrid, Spain \\
$^2$ Instituto Madrile\~no de Estudios Avanzados en Nanociencias (IMDEA-Nanociencia), 28049 Madrid, Spain\\
$^3$ Condensed Matter Physics Center (IFIMAC), Universidad Aut\'onoma de Madrid, 28049 Madrid, Spain.
}

\date{\today}

\begin{abstract}
High harmonic generation (HHG) from semiconductors and insulators has become a very active area of research due to its 
great potential for developing compact HHG devices. Here we show that by growing monolayers (ML) of insulators on 
single-crystal metal surfaces, one can tune the harmonic spectrum by 
just varying the thickness of the ultrathin layer, not the laser properties. This is shown from numerical solutions of 
the time-dependent Schr\"odinger equation for $n$ML NaCl/Cu(111) systems ($n=1-50$) based on realistic potentials 
available in the literature. Remarkably, the harmonic cutoff increases linearly with $n$ and as much as an order of magnitude when going from $n$ $=$ 1 to 30, while keeping the laser intensity low and the wavelength in the 
near-infrared range. Furthermore, the degree of control that can be achieved in this way is much higher than 
by varying the laser intensity. The origin of this behavior is the reduction of electronic ``friction'' when moving from the 
essentially discrete energy spectrum associated with a few-ML system to the continuous energy spectrum (bands) inherent 
to an extended periodic system.   
\end{abstract}

\pacs{42.65.Ky,78.66.Nk,42.65.-k}

\maketitle

Discovered in the 1980's~\cite{Mcpherson1987josab,Ferray1988jpb,Huillier1993prl}, high harmonic generation (HHG) has 
become the fundamental tool of modern attoscience~\cite{Agostini2004rpp,Scrinzi2006jpb,Krausz2009rmp}.  
In HHG from atomic or molecular gases, a strong laser field ionizes an electron, which then gains energy from the field and returns to the parent ion, 
where it finally recombines converting the gained energy into 
high-frequency radiation~\cite{Krause1992prl,Corkum1993prl,Schafer1993prl,Lewenstein1994pra}. The process repeats 
every 
half-cycle of the ionizing field, thus leading to a sequence of attosecond light bursts that contain multiples of the 
fundamental laser frequency, $\omega_0$. Since electronic motion in atomic and molecular systems occurs in the 
attosecond time scale, light pulses arising from HHG are currently used to probe electron dynamics in those systems 
~\cite{Drescher2002nature,Niikura2003nature,Uiberacker2007nature,Mauritsson2010prl,Goulielmakis2010nature,Wang2010prl,
Ott2014nature,Calegari2014science}. Furthermore the HHG process itself contains the signature of the parent-ion 
dynamics 
occurring during the round trip of the traveling electron. Thus the analysis of the HHG spectral features can also 
reveal important aspects of such dynamics 
~\cite{Lein2005prl,Baker2006science,Smirnova2009nature,Haessler2010natphys,Shiner2011natphys,Negro2014fardis,
Kraus2015science}.
 
HHG from condensed matter systems was first observed in the mid 90s~\cite{Linde1995pra,Norreys1996prl}. In these early 
experiments, bulk metals and dielectrics irradiated with very intense near-infrared laser fields (of the order of 
10$^{18}$ W/cm$^2$) were shown to emit harmonic radiation as a result of plasma oscillations induced in the system (see 
also~\cite{Dromey2006natphys})
Due to the high intensity of the field, HHG was 
always accompanied by sample damage. 
More recent experiments have made use of nano tips and nano spheres 
~\cite{Kim2008nature,Schenk2010prl,Kruger2011nature,Zherebtsov2011natphys,Ciappina2014pra}, from which HHG has been 
produced by using relatively moderate fields (of the order of 10$^{12}$ W/cm$^2$). HHG has 
also been observed in bulk semiconductors and insulators 
~\cite{Ghimire2011natphys,Zacks2012nature,Schubert2014natphot,Hohenleutner2015nature,Lun2015nature,Vampa2015nature,
Vampa2015prl} by using similar low intensities (even lower than those usually needed to generate high harmonics in the 
gas phase) and longer wavelengths, down to the mid infrared (MIR). Under these conditions, no significant damage of the 
sample is produced. One of the main conclusions of these experiments was that, at variance with HHG in the gas phase, 
the harmonic cut-off scales linearly with the applied field~\cite{Ghimire2011natphys,Vampa2014prl,Vampa2015prb}, which 
is the consequence of the electron moving in dispersive bands, usually the conduction bands 
(intraband dynamics~\cite{Pronin1994prb,Hawkins2013pra,Vampa2015prb}) and the 
electron tunneling through the various band gaps accessible in the system (interband dynamics 
~\cite{Plaja1992prb,Golde2008prb,Hawkins2015pra,Vampa2015prb,Catoire2015prl,Tamaya2016prl}). In spite of its apparent 
complexity, HHG in bulk semiconductors and insulators has become a very promising area of research, since, e.g., (i) it 
allows one to reconstruct the band structure of the system~\cite{Lun2015nature,Vampa2015nature,Vampa2015prl} when more
standard condensed-matter techniques
do not work, (ii) 
it is fairly robust against the presence of external fields~\cite{Vampa2015nature}, which is crucial for applications 
in 
electronics, (iii) it often leads to multi-plateau harmonic spectra~\cite{McDonald2015pra}, 
and more importantly (iv) it opens the way to the fabrication of compact HHG devices. 

In this context, it is worth exploring the possibility to 
modify and eventually control the high harmonic process by changing the properties of the material, not the laser 
characteristics. 
An appealing approach is to grow ultra thin layers of insulators (like NaCl, KCl, 
etc) 
on single-crystal metal surfaces. By just varying the thickness of the ultra thin layer, which is a standard procedure 
in surface physics, one would like to exert some control on the metal response, similar to that required to gradually 
modify the catalytic activity of metallic 
species~\cite{Rodriguez1996surfscirep,Gsell1998science,Creeley2004natmat,Otero2004surfsci,Laurent2008prb,Minniti2012jcp}
, 
to decouple molecules and self-assembled molecular networks from the metal substrate that holds them 
~\cite{Repp2005prl,Garnica2013natphys,Jarvinen2013nanolett,Robledo2015jpcc}, and, what is more important in the context 
of the present work, to gradually modify the response of metal surface electrons 
~\cite{Repp2001prl,Gross2009science,Diaz-Tendero2011prb,Barjenbruch1989surfsci,Kiguchi2005prb,Kim2014nanoconv,Tsay2009surfsci}. Inspired by the latter works, we 
propose a similar strategy to tune the harmonic cut-off. To ensure its feasibility, we choose a Cu(111) metallic 
substrate, which has been widely used to grow NaCl monolayers (ML) leading to $n$ML NaCl/Cu(111) composite systems 
~\cite{Repp2006apa,Gross2009science,Diaz-Tendero2011prb,Repp2001prl,Folsch2002prb,Bennewitz2006jpcm}. The Cu(111) 
surface 
has an additional advantage: it possesses a localized surface state that lies within the 
surface-projected band gap~\cite{Chulkov1999surfsci}.
If one considers irradiating the Cu(111) surface with linearly polarized light at grazing incidence, so that 
polarization goes along the (111) direction, a substantial number of surface electrons will efficiently scape from the 
metal surface and will be driven into the ultrathin NaCl layer every time the field points outwards 
the 
surface (i.e., once every laser cycle), thus avoiding screening and decoherence effects due to metal  
electrons (NaCl is nearly transparent to IR light and has a refraction index close to 1~\cite{Li1976jpcrd}). The 
idea 
is that only Cu electrons will move in the NaCl periodic potential, since NaCl electrons are tightly bound to the atomic 
centers and ionizing them with low intensity lasers and long wavelengths is much less probable. 

Here we show, by using realistic potentials for the $n$ML NaCl/Cu(111) systems, that the harmonic cutoff increases linearly with the number of NaCl monolayers and as 
much as an order of magnitude when going from $n$ $=$ 0 to 30. This is achieved by using rather 
low laser intensities, which is a necessary condition to avoid damaging the substrate, and wavelengths of the order of 
1-2 $\mu$m, i.e., not necessarily in the MIR region. The mechanisms of HHG emission from Cu(111) and 
$\infty$ML NaCl/Cu(111) are similar to those already stablished for gases and bulk semiconductors, respectively. 
However, the increase and strong variation of the HHG yield with the number of NaCl monolayers is due to a different 
mechanism: the decrease of electronic ``friction'' when moving from the essentially discrete energy spectrum 
associated with a few-ML system to the continuous energy spectra (bands) inherent to an extended periodic system. We 
show that the degree of control achieved in this way is much higher than by varying the laser intensity.     

\begin{figure}[t!]
\centering
\includegraphics[width=\columnwidth]{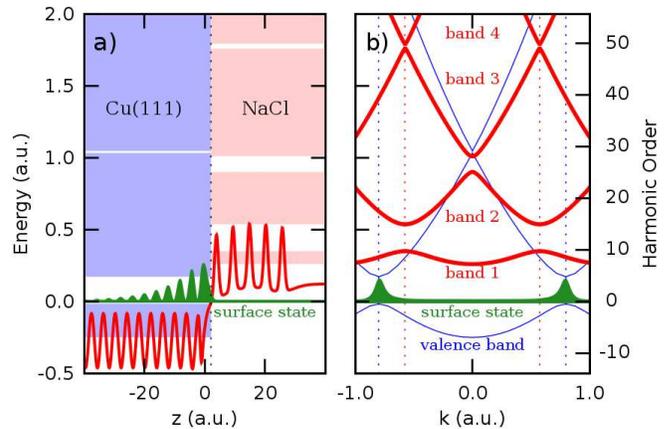}
\caption{
(a) Potential for the 5ML NaCl/Cu(111) system (red) and its surface state (green).
(b) Bands diagram for Cu(111) (blue) and NaCl (red). Vertical dashed lines indicate the limits of the 
corresponding first Brillouin zones. The momentum distribution of the surface state is also shown (green).
}
\label{fig: potential dos and bands}
\end{figure}

As in previous works \cite{Vampa2014prl,Vampa2015prb,Catoire2015prl}, we will make use of the single-active electron approximation. 
The evolution of the electronic quantum state $\psi(z,t)$ is governed by the 
time-dependent Schr\"{o}dinger equation (TDSE), $i\partial_t\psi(z,t)=H(p,z,t)\psi(z,t)$ (atomic units are used 
throughout unless otherwise stated), where, in the length gauge, the Hamiltonian is given by 
\begin{equation}
\begin{split}
H(z,p,t) = &\frac{1}{2}p^2 + zS(z,\zeta)F(t)E_0\cos(\omega_0t) \\
& + V(z) -i W_l(z-z_l) - i W_r(z-z_r).
\end{split}
\label{eq: hamiltonian}
\end{equation}
Here, $p$ is the electron momentum, $z$ the electron position referred to the Cu(111) image plane, $E_0$ the field amplitude,  
$\omega_0$ the laser frequency (we have chosen $\omega_0=0.03588$ a.u., i.e., wavelength $\lambda=1.27$ $\mu$m and optical period $T=2\pi/\omega_0 \approx 4.2$ fs), $F(t)$ is the pulse envelop defined as a two-cycle ramp-on $\sin^2$ function followed by a   
$10$-cycles flat-top segment and a two-cycle ramp-off $\sin^2$ function, $S(z,\zeta)$ is a function that accounts for screening of the electric field 
inside the Cu(111) surface 
\footnote{$S(z,\zeta) = \frac{1}{2}\Bigl\{1+\tanh\left[6\left(z+\zeta/2\right)/\zeta\right]\Bigr\}$}, 
$iW_l(z-z_l)$ and $iW_r(z-z_r)$ are complex absorbing potentials~\cite{Manolopoulos2002jcp} located at $z_l=-400.0$ and $z_r=500.0$ a.u. to avoid 
unphysical reflections on the box boundaries, and $V(z)$ is the potential that binds the electron to the system.
The screening function 
reduces the electric field to zero after a screening length $\zeta$. We have used $\zeta=4.0$ \AA (one) and $\zeta=7.6$ \AA (two atomic layers). Both lead to almost identical results. The $V(z)$ potential has been obtained from the lateral average of the accurate three-dimensional potential reported in \cite{Diaz-Tendero2011prb} and is given by 
\begin{equation}
\begin{split}
V(z) = &V_\text{Cu(111)}(z) + \sum_{i=1}^nV_\text{ion}(z-c_i)\\
& \qquad\qquad +V_\text{pol}(z-c_1)+V_\text{pol}(z-c_n),
\end{split}
\end{equation}
where $V_\text{Cu(111)}(z)$ is the Cu(111) model potential of Chulkov et al~\cite{Chulkov1999surfsci},
$V_\text{ion}(z-c_i)$ is a pseudopotential representing a NaCl atomic layer centered at $z=c_i$, and $V_\text{pol}(z-c_i)$ 
is an additional polarization potential induced by
the metal surface and the vacuum on the first and last monolayers, respectively. This potential reproduces fairly well the surface projected band gap 
of Cu(111), the energy of the corresponding surface state and the first three image states of the full dimensional system. Fig. \ref{fig: potential dos and bands}a shows the potential for the case of 5ML NaCl/Cu(111). As 
can be seen, the atomic positions in the NaCl region are associated with potential maxima, reflecting the fact that the 
interaction of an external (Cu) electron with the NaCl atomic centers is repulsive. 

The TDSE has been solved by using a modified version of the split-operator technique~\cite{Dion2014cpcomm}. We have 
used a spatial grid of equidistant points with $\Delta z$ = 0.1 a.u. in the interval $z \in [-500,600]$ a.u., and 
a constant time step of $0.01$ a.u. over the whole pulse duration. The harmonic 
yield has been obtained by Fourier transforming the time-dependent dipole operator calculated in the acceleration form.
To evaluate the initial surface state we have employed a filtering diagonalization method~\cite{Barinovs2002chemphys}. 
We note that this state is located below the Fermi level ($E_f=-0.180$ a.u.), just above the Cu valence band
(Fig. \ref{fig: potential dos and bands}a), and therefore it is occupied. Due to its localized nature, its 
representation in momentum space corresponds to two well defined peaks that appear in the limits of the first Brillouin 
zone of Cu (Fig. \ref{fig: potential dos and bands}b). In agreement with previous findings~\cite{Repp2004prl}, we have 
found that the presence of the NaCl layers barely affects this state.  

\begin{figure}[t]
\centering
\includegraphics[width=\columnwidth]{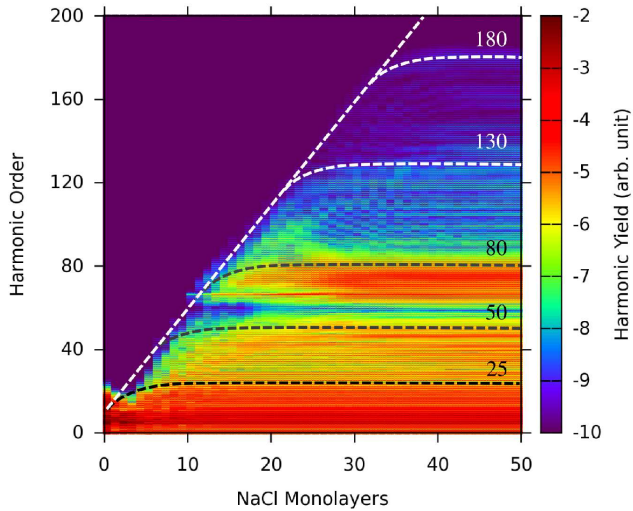}
\caption{
Harmonic spectra as a function of the number of NaCl monolayers. Yield is shown in logarithmic 
scale. The laser intensity is $20$ TW/cm$^2$. See text for details.
}
\label{fig: hhg NaCl effect}
\end{figure}

Fig.~\ref{fig: hhg NaCl effect} shows the calculated harmonic spectrum as a 
function of the number of NaCl layers, $n$, for a field amplitude/intensity of $0.024$ a.u./20 TW/cm$^2$. 
As can be seen, the harmonic cut-off increases monotonically with the number of NaCl monolayers, from ${\cal N}_c\sim 
10$ for 0 ML to ${\cal N}_c\sim 180$ for 35 ML, beyond which no further increase is observed. The cutoff scales almost 
linearly with $n$ and follows the approximate rule ${\cal N}_c = 5n$. One can also see the appearance of 
multi-plateau structures, whose number also depends on $n$. The different plateau limits are indicated by horizontal 
dashed lines. The first plateau extends up to the 25th harmonic and is only observed for $n\geq 5$. 
This plateau is followed by other ones extending up to the 50th, 80th, 130th and 180th harmonics for 
$n\geq 10, 15,20,30$, respectively. 
As can be seen from Fig. \ref{fig: potential dos and bands}b, the plateau appearance thresholds approximately correspond 
to the bottom of the NaCl bands in which the ejected electron moves, thus suggesting a steplike process that populates 
the successive NaCl bands: harmonic orders below ${\cal N}_c = 25$ would reflect population of the second band, those between 25 
and 50, that of the third band, and so on. This would be similar to the interband mechanism described in previous work 
(see \cite{Catoire2015prl} and references therein). Interestingly, the present results show that the efficiency of such mechanism increases with the number of overlayers: an increase in $n$ leads to a proportional increase in 
the intraband density of states (DOS), thus to less ``friction" during the intraband motion, and hence to a linear increase of the 
cut-off with $n$. The rate of this increase, $\sim$5 for $n$ML NaCl/Cu(111), is dictated by the width and height of the potential energy barriers (i.e., by the width of the bands and the gaps) in the $n$ML region, thus it can be tuned by using insulators with different lattice constants, e.g. LiF, KCl, etc. 

\begin{figure*}[t]
\includegraphics[width=\textwidth]{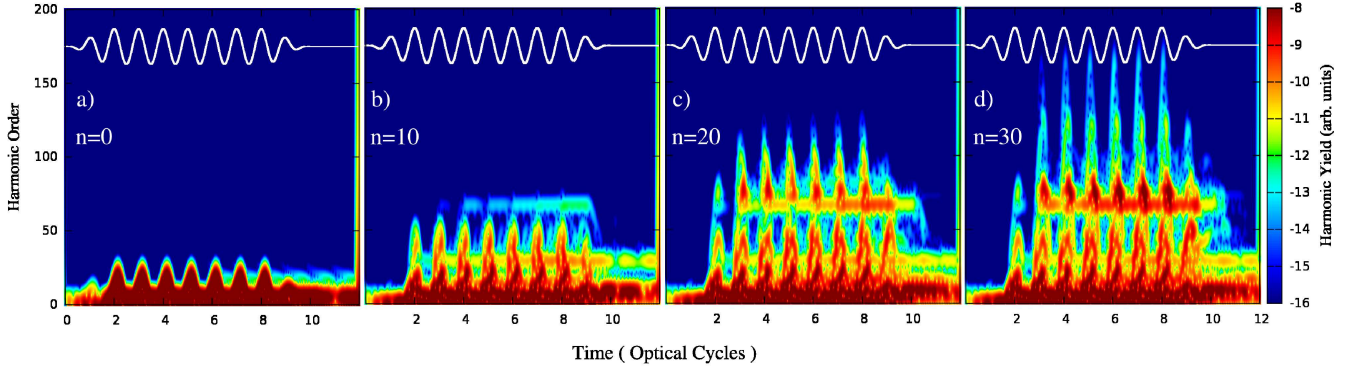}
\caption{Time-windowed Fourier transform of the harmonic spectra for $n$ML NaCl/Cu(111). The laser intensity is 20 TW/cm$^2$.}
\label{fig: gabor profiles}
\end{figure*}

The above mechanism is further confirmed by the windowed Fourier transforms of the time dependent dipole (Gabor 
profiles~\cite{Gabor1946jiee}) shown in Fig.~\ref{fig: gabor profiles} for the two extreme cases, clean 
Cu(111) and 30ML NaCl/Cu(111), and the 10 and 20 ML NaCl/Cu(111) intermediate cases.
As in the atomic case, high-order harmonics are associated with long 
Gabor trajectories, which show up when the field reaches its maximum intensity, i.e., after the third cycle of the 
current pulse. However, the trajectories are seen once every laser cycle instead of once every half cycle, in contrast with atomic systems. This is due to screening of the laser field by the Cu electrons, which suppresses all 
trajectories that start at the Cu(111) surface and then go into the metal (negative field), while  
trajectories going towards the NaCl layers survive (positive field). For Cu(111) (Fig. ~\ref{fig: gabor profiles}a) the 
trajectory profiles are very similar to those found in atoms. The situation is completely different in the presence of 
the NaCl monolayers (Fig.~\ref{fig: gabor profiles}(b-d)): trajectories are longer and longer and involve more and more 
bands as $n$ increases. As can be seen, electrons that reach a particular band can either return within the same band or 
overcome the gap and jump to a higher band, where they go on acquiring kinetic energy before returning. The jumps 
between bands lead to depletions in the Gabor intensity, since as soon as the electron reaches a new 
band, the electron can more easily progress within this new band than go back to the previous band, which requires going through the gap again. 
It is worth noticing that harmonic emission from pure NaCl would not 
be so efficient and would not lead to such extended plateaus as for $n$ML NaCl/Cu(111) ($n>30$), because the band gap 
between the valence and the conduction bands in bulk NaCl is much larger than the gap between the Cu(111) surface state 
and the lowest conduction band in $n$ML NaCl/Cu(111) ($8.5$-$9.0$ eV~\cite{Roessler1968pr,Poole1975prb} vs. $5.1$ eV, 
respectively).

\begin{figure}[b]
\includegraphics[width=1.0\columnwidth]{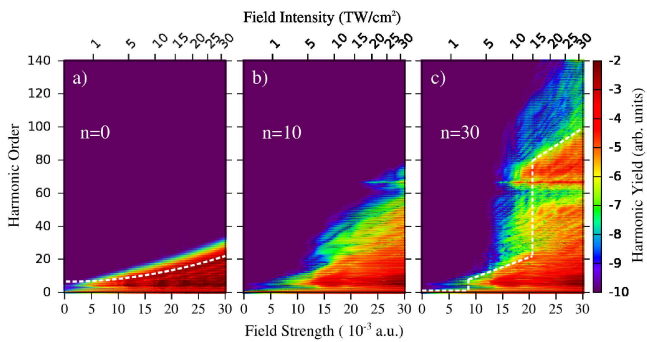}
\caption{Harmonic  yield  as  a  function  of  laser  field  intensity/strength for $n$ML NaCl/Cu(111).
Yield  is  shown  in logarithmic scale. See text for details. }
\label{fig: yield vs field strength}
\end{figure}

Figure \ref{fig: yield vs field strength} shows the variation of the harmonic yield with the field amplitude $E_0$ for 
$n$ $=$ 0, 10, and 30. For the clean Cu(111) surface, the harmonic cutoff increases with the square of the field, in 
agreement with the prediction of the three-step model~\cite{Corkum1993prl}, ${\cal N}_c=W_s + 3.17E_0^2/4\omega_0^2$ (dashed line in 
Fig. \ref{fig: yield vs field strength}a), 
where $W_s=0.196$ a.u. is the energy required to excite the surface electron to the Cu conduction band. This is not surprising because the round trip of the electron, initially 
localized on the surface, occurs in vacuum. The difference in the present case is that such trip is only possible 
in half of the accessible space. 

The behavior is much more complex in the presence of NaCl monolayers. For the 30ML NaCl/Cu(111) system
(Fig. \ref{fig: yield vs field strength}c), there are regions where the cutoff increases linearly with $E_0$ and others 
where it suddenly increases at specific field values (dashed lines). The first of these sudden increases occurs at 
$E_0^{(1)}=0.008$ a.u.. This is the value of the field required to, first overcome the gap between the Cu(111) surface 
state and the lowest NaCl, then promote the electron to the top of the latter band, and finally overcome the gap between 
the first and the second NaCl bands (see Fig.~\ref{fig: potential dos and bands}b). This effect leads to an extension 
of the cutoff of about 10 harmonic orders. After this, the cutoff increases almost linearly with $E_0$ until a second, 
even more sudden increase occurs at $E^{(2)}_0=0.021$ a.u.. This is the field needed to additionally promote the electron to the top of the second NaCl band and overcome
the gap between this and the third band. As soon as the electron reaches the third NaCl band, there are practically no more gaps to overcome and the electron can easily progress through the band series by means of one-photon resonant transitions near the Brillouin zone, in a way similar to that reported in \cite{Catoire2015prl} 
to explain strong field ionization from periodic systems. This effect leads to an 
extension of the cutoff by around 100 orders. 

As shown by Vampa et al.~\cite{Vampa2015prb}, linear scaling with $E_0$ is the consequence of the electron 
traveling through energy dispersion bands. The cutoff law resulting from this interband processes approximately follows 
the expression ${\cal N}_c^{ij}=[\varepsilon^{ij}(\kappa)-\kappa\,d_k\varepsilon^{ij}(\kappa)] + 
d_k\varepsilon^{ij}(\kappa)E_0\delta/\omega_0 $ where, in our case, $i$ and $j$ refer to two NaCl bands, 
$\varepsilon^{ij}(k)$ and $d_k\varepsilon^{ij}(k)$ are the corresponding band gaps and $k$ derivatives, respectively, 
$\kappa$ is the point in reciprocal space where $\varepsilon^{ij}(k)$ varies almost linearly with $k$ and $\delta=1.23$. 
Fig. \ref{fig: yield vs field strength}c (dashed lines) shows linear cutoff laws derived from this formula for 
electron-hole pair re-collisions coming from bands (2,1) ($0.008$ a.u. $< E_0 < 0.021$ a.u.) and (4,1) ($E_0 > 0.021$ 
a.u.). For the intermediate 10ML NaCl/Cu(111) system (Fig. \ref{fig: yield vs field strength}b), the 
situation is more subtle. As discussed above, a reduction in 
the number of NaCl monolayers leads to a reduction of the DOS within every band and, consequently, to less extended harmonic spectra. In general, the degree of control that can 
be exerted by varying the number of layers is much superior than by varying the field strength, since, in the first 
case, variations of the harmonic spectrum are linearly monotonic (Fig. \ref{fig: hhg NaCl effect}), while, in the second, 
they are abrupt and mainly occur at a few selected values of the field. 

In conclusion, our solutions of the TDSE for $n$ML NaCl/Cu(111) systems by using realistic potentials have allowed us to 
show that harmonic emission plateaus can be extended by more than an order of magnitude by varying the number of NaCl 
monolayers from 0 to 30. The cutoff extension scales linearly with the number of layers as a consequence of the linear increase of the intraband density of states. Thus, the degree of control that can be achieved in this way is much higher than that obtained by varying the laser intensity. 

We gratefully acknowledge Dr. Sergio D\'{\i}az-Tendero for his valuable advice on the construction of the surface 
potentials. We also acknowledge allocation of computer time at Mare Nostrum BSC and CCC-UAM. This work was supported by 
the European Research Council Advanced Grant No. XCHEM 290853, the MINECO Project No. FIS2013-42002-R, the European COST 
Action XLIC CM1204, and the CAM project NANOFRONTMAG.

\end{document}